\documentclass[twocolumn,showpacs,prb,aps,epsf,amsmath,amssymb]{revtex4}
\usepackage{graphicx}
\usepackage{amsbsy}
\newlength{\onewidth}
\usepackage{tensor}
\usepackage{multirow}
\usepackage{epsfig}
\begin{document}

\title{Thermal Conductivity of InAs/GaSb Superlattice}

\author{Chuanle~Zhou}
\affiliation{
Electrical Engineering and Computer Science, Northwestern University, Evanston, IL 60208 USA
}
\author{B.-M. Nguyen}
\affiliation{
Electrical Engineering and Computer Science, Center for Quantum Devices, Northwestern University, Evanston, IL 60208 USA
}
\author{M.~Razeghi}
\affiliation{
Electrical Engineering and Computer Science, Center for Quantum Devices, Northwestern University, Evanston, IL 60208 USA
}
\author{M.~Grayson\footnote{Corresponding author: m-grayson@northwestern.edu}}
\affiliation{
Electrical Engineering and Computer Science, Northwestern University, Evanston, IL 60208 USA
}

\begin{abstract}
The cross-plane thermal conductivity of a type II InAs/GaSb superlattice (T2SL) is measured from 13~K to 300~K using the 3$\omega$ method. Thermal conductivity is reduced by up to 2 orders of magnitude relative to the GaSb bulk substrate.  The low thermal conductivity of around 1-8~W/m$\cdot$K may serve as an advantage for thermoelectric applications at low temperatures, while presenting a challenge for T2SL quantum cascade lasers and high power light emitting diodes.  We introduce a power-law approximation to model non-linearities in the thermal conductivity, resulting in increased or decreased peak temperature for negative or positive exponents, respectively.

\end{abstract}


\maketitle


InAs/GaSb type II broken-gap superlattices (T2SL) have been successfully developed for use in long wavelength (LWIR) and midwave (MWIR)  infrared detectors\cite{Smith}$^-$\cite{Nguyen07} with band engineered cut-off wavelengths and a reduction in dark current compared to direct gap bulk semiconductors.  High-power LWIR and MWIR cascaded LEDs and lasers based on T2SL band structure engineering\cite{Yang96,Koerperick09,Koerperick11,Rejeb}  also offer tunable infrared light emission. Both detectors and lasers perform best at low temperatures,\cite{Abdollahi} and heat needs to be carried away from the devices to reduce Auger recombination and increase carrier lifetime for detectors, and to decrease lasing threshold for lasers. It is therefore important to have a good understanding of the thermal conductivity of this artificial material, to understand performance parameters at low operating temperatures.  Recently, efforts by the authors have also identified the T2SL as a candidate material for low-temperature Peltier cooling\cite{Zhou} based on the large Seebeck coefficients that have been reported,\cite{Cao} but measurements of the thermal conductivity at cryogenic temperatures were lacking. In this work, we measure the cross-plane thermal conductivity of T2SLs from 13~K to 300~K using the 3$\omega$ method,\cite{Cahill90,Cahill94} and present a power-law approximation for modeling thermal conductivity over large thermal gradients expected to occur in such low-thermal conducting materials.


The two different T2SLs studied in this work were grown by molecular beam epitaxy on GaSb substrates. Because this material was intended for p-i-n detectors, it consists of a 0.5 $\mu$m GaSb p$^+$ ($\sim10^{18}$ cm$^{-3}$) buffer layer, followed by a 0.5 $\mu$m T2SL p$^+$ ($\sim10^{18}$ cm$^{-3}$) region, a 2 $\mu$m undoped T2SL layer, a 0.5 $\mu$m T2SL n$^+$ ($\sim10^{18}$ cm$^{-3}$) region, and a 10 nm Si doped InAs n$^+$ capping layer.  T2SL-1 is composed of 12 monolayers (ML) of InAs and 8 ML of GaSb per period. T2SL-2 is composed of 19 ML of InAs and 18 ML of GaSb per period.\cite{protocol}
\begin{figure}
	\includegraphics[width=\columnwidth]{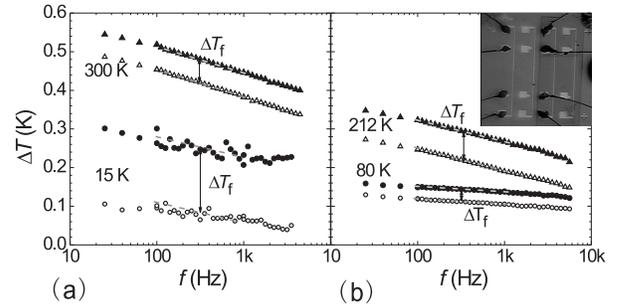}
    \caption{Temperature change $\Delta T$ vs. frequency $f$ measured with the 3$\omega$ method for the T2SL + SiO$_2$ + GaSb substrate (filled symbols) and for SiO$_2$ + GaSb substrate (open symbols), (a) at 15~K and 300~K, and (b) at 80~K and 212~K. Inset: picture of the sample, with left half etched to substrate, and right half unetched. Two Au heater-thermometer lines are deposited on each side, with current contact pads at top/bottom and voltage contacts in between.}
	\label{fig:dT}
\end{figure}

Following standard 3$\omega$ sample preparation, the thermal conductivity of a sample including the layer of interest is measured relative to a reference sample without the layer.  Thus the T2SL is wet-etched away from half the sample with a solution of citric acid 
and phosphoric acid plus peroxide. 
An insulating SiO$_2$ layer of $150$~nm is deposited using plasma enhanced chemical vapor deposition (PECVD) to prevent an electrical short circuit through the conducting substrate. Then 200~nm thick gold heater-thermometer filaments are deposited atop a $3$~nm Ti adhesion layer using e-beam evaporation on both etched and unetched regions. The filament shown in the inset of Fig.~\ref{fig:dT}(b) is 3.6~mm long and 30 $\mu$m wide, much wider than the 3~$\mu$m T2SL thickness so that the heat flow through the T2SL obeys the one-dimensional thermal diffusion equation. \cite{Cahill94}

According to the 3$\omega$ method,\cite{Cahill90,Cahill94} by sending a current excitation at frequency $\omega$ through the gold filament, heating power $P$ is induced at frequency $2\omega$, leading to thermal variation $\Delta T$ at the same frequency and ultimately a 4-point voltage $V_{3\omega}$ at frequency $3\omega$ proportional to the temperature difference via the thermal coefficient of Au.\cite{Cahill90}  The thermal difference $\Delta T = \Delta T_\mathrm{sub}(\omega) + \Delta T_\mathrm{f}$ can have a frequency dependent substrate contribution and a frequency independent thin-film contribution,
\begin{equation}
\Delta T_\mathrm{sub} (\omega)= \frac{-P}{l \pi \kappa_\mathrm{sub}}  \frac{1}{2} \ln (2\omega)   + C
\end{equation}
\begin{equation}
\Delta T_\mathrm{f}  = \frac{P}{\kappa_\mathrm{f}}\frac{t}{lw}
\end{equation}
where $\kappa_\mathrm{sub}$ is the thermal conductivity of the GaSb substrate, $w$ and $l$ the width and length of the Au filament heater, and $C$ is a constant offset which includes the SiO$_2$ layer contribution.  For the T2SL film, $\kappa_\mathrm{f}$ and $t$ are the film thermal conductivity and thickness, respectively.

We measured the sample in an Oxford variable temperature insert (VTI) helium gas flow cryostat from 300~K down to 13~K, using standard lock-in techniques. 
Since there are background 3$\omega$ voltages from the lock-in power source and nonlinear components in the measurement circuit and in the lock-in A-B input channels, we also measured a reference background 3$\omega$ signal with a low thermal coefficient resistor of equal resistance  as the gold filament. 

\begin{figure}
	\includegraphics[width=\columnwidth]{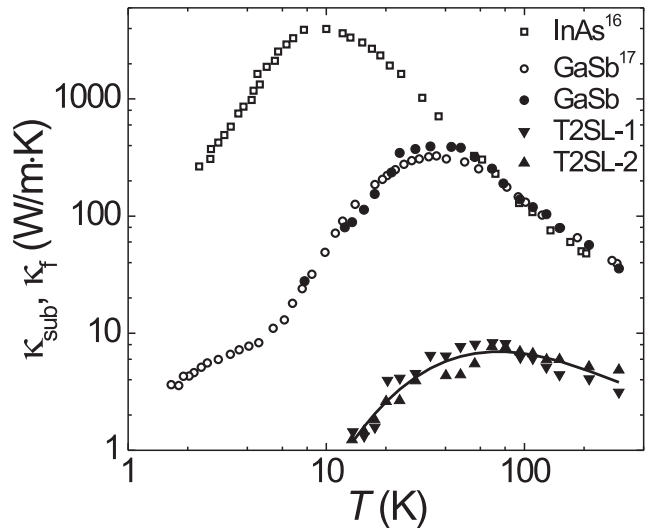}
    \caption{Cross-plane thermal conductivity for T2SL samples (solid triangles). Measured GaSb substrate thermal conductivity is shown in solid circles. Published data for bulk thermal conductivity of InAs \cite{Tamarin} and GaSb \cite{Holland} are shown for comparison in open squares and open circles respectively. The solid line is a fitting polynomial for the average value of T2SL thermal conductivity described in the text.}
	\label{fig:SL_k}
\end{figure}

By measuring the slope of the temperature difference $\Delta T$ as a function of log-frequency \cite{Cahill90}, we can deduce the substrate thermal conductivity and thereby confirm the reliability of our measurements. In Fig.~\ref{fig:SL_k}, we compare our measured GaSb substrate thermal conductivity (solid circles) with previously published GaSb bulk thermal conductivity (open circles)\cite{Tamarin,Holland}, indicating excellent agreement.

The T2SL cross-plane thermal conductivity $\kappa _{\mathrm f}$  is deduced from the thermal difference $\Delta T_\mathrm{f}$, 
plotted with solid triangles in Fig.~\ref{fig:SL_k}.  As expected from other studies of superlattice thermal conductivities, the T2SL value is reduced by 2 orders of magnitude compared with the bulk substrate thermal conductivities for GaSb bulk (open circles).  We note that the suppression is much greater, up to as much as 3 1/2 orders of magnitude, when compared with InAs bulk (open squares).
%
To estimate the T2SL thermal conductivity at arbitrary temperatures, we provide the equation for the solid-line empirical fit to the measured T2SL thermal conductivity in Fig.~\ref{fig:SL_k}.
\begin{equation}
\label{eq:logkappa}
\log \kappa = \log \kappa_{m}-B\left[ \log\left(\frac{T}{T_m}\right)\right]^2+C\left[\log\left(\frac{T}{T_m}\right)\right]^3.
\end{equation}
with maximum thermal conductivity $\kappa_{m} = 6.954$~W/m$\cdot$K at $T_m$ = 74~K, $B = 1.0312$, and $C = 0.53042$.

To improve device performance, it is useful to model thermal distribution within the active layer of a quantum cascade laser (QCL) or LED \cite{Spagnolo,Howard}. 
With poor thermal conductivity, one can expect large thermal gradients across the T2SL layer, so it is useful to develop an analytical estimate of the thermal profile including large temperature drops.  We do so below by introducing a power-law approximation to the temperature dependence of the thermal conductivity. 

The thermal profile can be determined from the power density generated per unit volume for the various devices of interest, namely LEDs and QCLs.  For an infrared LED the power density due to Joule heating is
\begin{equation}
\label{eq:P_LED}
P = {\cal E}J~,
\end{equation}
where the electric field $\cal E$ across the T2SL emitting layer is assumed uniform, $J$ is the current density and the light output power is neglected. For continuous-wave (CW) lasers, one can estimate the heat dissipated at threshold assuming that almost all input power is dissipated as heat
\begin{equation}
\label{eq:P_laser}
P = (E_{32}+\Delta)J_{th}/eL_p~,
\end{equation}
where $J_{th}$ is the threshold current density, $E_{32}$ is the energy of the optical transition, $\Delta$ is the energy separation between the ground state of the injector to the lower laser level of the previous active region, and $L_p$ is the length of one period of QCL active region and injector.\cite{Howard}.  

Using Eqs.~\eqref{eq:P_LED}-\eqref{eq:P_laser}, the thermal profile in the active region becomes:
\begin{equation}
\label{eq:power}
P = \nabla \cdot (-\kappa \nabla T).
\end{equation}
For a substrate-side mounted device, the primary heat dissipation is through the GaSb substrate, modeled as a 1D thermal diffusion problem  \cite{Spagnolo}. If, for the moment, a constant thermal conductivity $\kappa_0$ is assumed, the solution to Eq.~\eqref{eq:power} takes the simple form\cite{Edwin}
\begin{equation}
\label{eq:dT(x)}
\Delta T(x) = - \frac{1}{2} \frac{P}{\kappa_0}x^2 + \frac{dT}{dx}\bigg|_0 x.
\end{equation}
with boundary conditions $\Delta T(0) = 0$ and $dT/dx(0) = dT/dx|_0$.

\begin{figure}
	\includegraphics[width=\columnwidth]{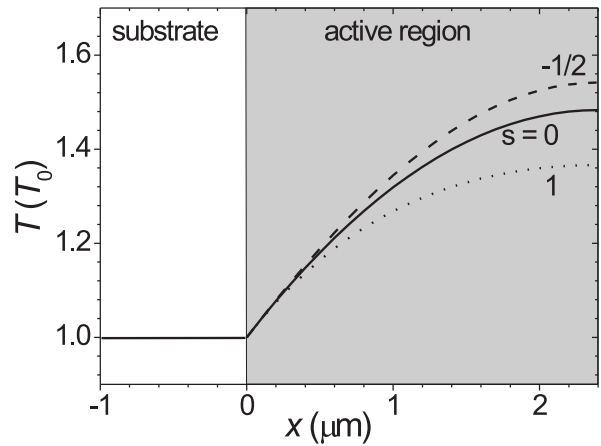}
    \caption{Calculated temperature profile in the active region normalized to the cold-sink substrate temperature $T_0$  with a typical power dissipation in a T2SL LED. The power-law exponent $s=0$ approximates the behavior in the intermediate temperature range $T_0 = 50-100$~K whereas the $s = -1/2$ negative exponent is appropriate to higher temperatures $T_0 = 150-300$~K and $s = 1$ to lower temperatures $T_0 = 13-30$~K.}
	\label{fig:dT-x}
\end{figure}

However, we arrive at a more accurate estimate of the thermal profile at high powers if we include the temperature dependence of $\kappa(T)$. By inspection of the T2SL log-log plot in Fig.~\ref{fig:SL_k}, for large ranges of temperature above 150~K and below 30~K the slope is roughly constant and $\kappa(T)$ can be approximated with a power-law.  We thus empirically fit the T2SL thermal conductivity with the local power-law expression $\kappa(T) = \kappa_0 (\frac{T}{T_0})^s$, where $s$ is the power-law exponent, $T_0$ is the base-line temperature of interest, and $\kappa_0$ is the thermal conductivity at that temperature. Solving Eq.~\eqref{eq:power} under this local power-law assumption, the exact solution for the temperature profile in the active T2SL region with the same boundary conditions becomes:
\begin{equation}
\label{eq:dT-x_exact}
T(x) = T_0 \left[ {1+ (s+1)\frac{\Delta T(x)}{T_0} }\right] ^{\frac{1}{s+1}}
\end{equation}
where $\Delta T(x)$ is the same expression from Eq.~\eqref{eq:dT(x)} and $s$ is the power-law exponent.  In real devices, the change in temperature $\Delta T(x)$ will normally not exceed the absolute base temperature $T_0$, so this expression can be expanded for small $\Delta T(x) / T_0 < 1$
\begin{equation}
\label{eq:dT-x}
T(x) \simeq T_0 + \Delta T(x) - \frac{1}{2} s \frac{\Delta T^2(x)}{T_0}~.
\end{equation}
The first two terms describe the thermal distribution for constant baseline thermal conductivity $\kappa_0$, and the third term accounts for the local power-law assumption, proportional to the exponent $s$.  Note that the position of maximum temperature does not change between Eqs.~\eqref{eq:dT(x)} and \eqref{eq:dT-x}, nor does the thermal derivative boundary condition at $x=0$.
The resulting thermal distribution is plotted in Fig.~\ref{fig:dT-x} under high power dissipation $P$ such that the maximum temperature is about a factor of 1.5 greater than the substrate absolute temperature $T_0$.  When the exponent is positive the peak temperature is reduced, for example $s \sim 1$ for $T_0$ below 30~K.  And when the exponent negative the peak temperature is increased, for example $s \sim -1/2$ for $T_0$ above 150~K.

The local power-law exponent $s$ in the neighborhood of an operating temperature $T_0$ can be determined from the same empirical fit parameters as for Eq.~\eqref{eq:logkappa}:
\begin{eqnarray}
\label{eq:s}
s(T_0) & = & \frac{d \log\kappa}{d \log T}\bigg| _{T_0}
\nonumber\\
& = & -2B\log\left(\frac{T_0}{T_m}\right)+3C\left[\log\left(\frac{T_0}{T_m}\right)\right]^2.
\end{eqnarray}

We remark that these materials could be promising as cryogenic thermoelectrics because below 20~K the thermal conductivity is quite low of order 1~W/m$\cdot$K, and because high Seebeck coefficients up to 2 mV/K have been reported for the hole band and 300 $\mu$V/K for the electron band\cite{Cao} at 4~K. 
We thus conclude that careful thermal modeling with the knowledge of the thermal conductivity reported here can improve device performance for a range of applications.

{\em Acknowledgement - }
This work is supported by AFOSR grant FA-9550-09-1-0237 and NSF MRSEC grant DMR 0520513 through both an instrumentation grant and an NSF MRSEC Fellowship.


%

\end{document}